\documentclass[aps,prd,superscriptaddress,showpacs,showkeys]{revtex4}
\usepackage{amsmath}
\usepackage{amssymb}
\usepackage{palatino}
\usepackage{amsmath,tabstackengine}
\usepackage[colorinlistoftodos]{todonotes}
\usepackage[colorlinks=true,citecolor=black,linkcolor=red,anchorcolor=green,anchorcolor=green,urlcolor=blue]{hyperref}
\usepackage[T1]{fontenc}
\usepackage{graphicx}
\usepackage[english]{babel}
\usepackage{amsfonts}
\newcommand{\beq}{\begin{equation}}
\newcommand{\beqn}{\begin{eqnarray}}
\newcommand{\eeq}{\end{equation}}
\newcommand{\eeqn}{\end{eqnarray}}

\usepackage{MnSymbol,wasysym}
\usepackage{braket}
\usepackage{eurosym}
\usepackage{calrsfs}

\usepackage{changes}

\numberwithin{equation}{section}

\begin{document}

\title{Tunneling Glashow-Weinberg-Salam model particles from Black Hole
Solutions in Rastall
Theory}

\author{Ali \"{O}vg\"{u}n} 
\email{ali.ovgun@pucv.cl}
\affiliation{Instituto de F\'{\i}sica, Pontificia Universidad Cat\'olica de
Valpara\'{\i}so, Casilla 4950, Valpara\'{\i}so, Chile}

\affiliation{Physics Department, Arts and Sciences Faculty, Eastern Mediterranean University, Famagusta, North Cyprus via Mersin 10, Turkey}

\author{Wajiha Javed}

\email{wajiha.javed@ue.edu.pk} 

\author{Riasat Ali}
\email{riasatyasin@gmail.com}

\affiliation{Division of Science and Technology, University of Education, Township Campus, Lahore, Pakistan}
\date{\today }

\begin{abstract}
Using the semiclassical WKB approximation and Hamilton-Jacobi method, we solve an equation of motion for the Glashow-Weinberg-Salam model, which is important for understanding the unified gauge-theory of weak and electromagnetic interactions. We calculate the tunneling rate of the massive charged W-bosons in a background of electromagnetic field to investigate the Hawking temperature of black holes surrounded by perfect fluid in Rastall theory. Then, we study the quantum gravity effects on the generalized Proca equation with generalized uncertainty principle (GUP) on this background. We show that quantum gravity effects leave the remnants on the Hawking temperature and the Hawking radiation becomes nonthermal.

\end{abstract}

\maketitle

\section{Introduction}

\label{Introduction}
General relativity is analogously linked to the thermodynamics and quantum effects which strongly support it \cite{man1,man2,man3}. Black holes are the strangest objects in the Universe and they arise in general relativity, a classical theory of gravity, but it is needed to include quantum effects to understand the nature of the black holes properly. After Bekenstein found a relation between the surface area and entropy of a black hole \cite{Bekenstein:1974ax}, Hawking theoretically showed that black holes with the surface gravity $\kappa$ radiate at temperature $\kappa/2 \pi$ \cite{Hawking:1974rv,Hawking:1974sw}. On the other hand, Bekenstein-Hawking radiation causes the information loss paradox because of the thermal evaporation. To solve the information paradox, recently soft-hair idea has been proposed by Hawking et al. \cite{Hawking:2016msc}.

Since Bekenstein and Hawking great contribution on the black hole's thermodynamics, the radiation from the black hole is gained attention from researchers. There are many different methods to obtain the Bekenstein-Hawking radiation using the quantum field theory or the semiclassical methods. The quantum tunneling method is one of them \cite{Parikh:1999mf,Kerner:2006vu,Kerner:2007jk,Banerjee:2008cf,Akhmedov:2006pg,Akhmedov:2006un,Akhmedova:2010zz,Akhmedova1}. Nozari and Mehdipour \cite{c} have studied the Hawking radiation as tunneling phenomenon for Schwarzschild BH in noncommutative spacetime.
Nozari and Saghafi \cite{d} have investigated the tunneling of massless particles
for Schwarzschild BH by considering quantum gravity effects. The semiclassical tunneling method by using the  Hamilton-Jacobi ansatz with WKB approximation is another way to obtain the Bekenstein-Hawking temperature and the tunneling rate as $\Gamma \approx exp[-2 Im S]$ \cite{Angheben}. Different kind of particles such as bosons, fermions and vector particles are used to study the tunneling of the particles from the black holes and wormholes and obtain their Hawking temperature \cite{Kruglov1,Kruglov2,Kuang:2017sqa,Sakalli:2017ewb,Jusufi:2017trn,Sakalli:2016mnk,Ovgun:2016roz,Sakalli:2016cbo,Ovgun:2015box,Sakalli:2015jaa,Sakalli:2015taa,aji, Sakalli:2016fif,  Singh:2017car,Pu:2017sio,Ali:2017org,Javed:2017cok,Javed:2017saf,Jusufi:2016fpg,Jusufi:2016nxo,Ali:2013oja,Ali:2007xz,Ali:2007sh,  Sakalli:2012zy,Sakalli:2010yy,Pasaoglu:2009te,Sakalli:2015nza}. 
Nozari and Sefidgar \cite{e} have discussed quantum corrections approach to study BH thermodynamics. Nozari and Etemadi \cite{f} have investigated the KMM seminal work in case of a maximal test particles momentum. They showed that in the presence of both minimal length and maximal momentum there is no divergence in energy spectrum of
a test particle. Moreover, the uncertainty principle is modified as a generalized uncertainty principle (GUP) \cite{kempf,gup} to work on the the effect of the quantum gravity which is applied on the different areas. The important contribution of the GUP is to remove the divergences in physics. On the other hand, GUP can be used to modify Proca equation and Klein-Gordon equation to obtain the effects of the GUP on the Hawking temperature and check if it leaves remnants \cite{Tawfik:2015kga,Anacleto:2015mma,Dehghani:2015zca,Ali:2015zua}.
Nozari and Mehdipour \cite{g} have discussed BH remnants and their
cosmological constraints. Moreover, GUP is used to modify the thermodynamics of N-dimensional
Schwarzschild-Tangherlini black hole, speed of graviton, and the Entropic Force.
Feng et al. \cite{a} have studied the difference between the
propagation speed of gravitons and the speed of light by using GUP. They have also
investigated the modified speed of graviton by considering GUP.
Rama \cite{b} have studied the consequences of GUP which leads to varying speed of light and modified dispersion relations, which are likely to have implications for cosmology and black hole physics.

Since Maxwell, it was the dream of theoretical physicist to unify the fundamental forces in the nature in a single equation.
Glashow, Weinberg and Salam unified the theory of weak and electromagnetic interactions as an electroweak interaction in the 1960s. They assumed that the symmetry between the two different interactions would be clear at very large
momentum transfers. However, at low energy, there is a mass difference between the photon and the $W_+$, $W_-$, and $Z_0$ bosons which break the symmetry.


This paper is organized as follows: In Section 2, we investigate the Hawking temperature of the black hole solutions surrounded by perfect fluid in Rastall
theory using the tunneling of the massive vector particles. For this purpose we solve the equation of the motion of the Glashow-Weinberg-Salam model using the semiclassical WKB approximation with Hamilton-Jacobi method. In Section 3. we use the GUP-corrected Proca equation to investigate the tunneling of massive uncharged vector particles for finding the corrected Hawking temperature of the black hole
solutions surrounded by perfect fluid in Rastall theory. In Section 4,
we conclude the paper with our results.

\section{Tunneling of Charged Massive Vector Bosons}
In this section, we study the tunneling of the charged massive bosons from the different types of black holes surrounded by the perfect fluids in Rastall theory. 
\subsection{The Black Hole Surrounded by the Dust Field in Rastall theory}
First we study the line element of the black hole surrounded by the dust field \cite{R1}
\begin{eqnarray}
ds^{2}&=&-(1-\frac{2M}{r}+\frac{Q^{2}}{r^{2}}-
\frac{N_{d}}{r^{\frac{1-6\kappa\lambda}{1-3\kappa\lambda}}})dt^{2}
+\frac{dr^{2}}{1-\frac{2M}{r}+\frac{Q^{2}}{r^{2}}-
\frac{N_{d}}{r^{\frac{1-6\kappa\lambda}{1-3\kappa\lambda}}}}\nonumber\\
&&+r^{2}(d\theta^{2}+\sin^{2}\theta d\phi^{2}),\label{a11}
\end{eqnarray}
where $M$, is a mass of black hole, $\kappa$ and $\lambda$ are the
Rastall geometric parameters, $N_{d}$ dust field structure parameter
and $Q$ is a charge of black hole. Now, we can rewrite the
Eq.(\ref{a11}) in the following form
\begin{equation}
ds^{2}=-\added{G}(r)dt^{2}+B(r)dr^{2}+C(r)d\theta^{2}+D(r)d\phi^{2},
\end{equation}
where $\added{G}(r)$, $B(r)$, $C(r)$ and $D(r)$ are given below:
\begin{eqnarray*}
\added{G}(r)&=&1-\frac{2M}{r}+\frac{Q^{2}}{r^{2}}-
\frac{N_{d}}{r^{\frac{1-6\kappa\lambda}{1-3\kappa\lambda}}},~~
C(r)=r^{2},\\B(r)&=&\frac{1}{1-\frac{2M}{r}+\frac{Q^{2}}{r^{2}}-
\frac{N_{d}}{r^{\frac{1-6\kappa\lambda}{1-3\kappa\lambda}}}},~~
D(r)=r^{2}\sin^2\theta,
\end{eqnarray*}

The equation of motion for the Glashow-Weinberg-Salam
model \cite{R2,we,we2} is
\begin{equation}
\frac{1}{\sqrt{-\textbf{g}}}\partial_{\mu}(\sqrt{-\textbf{g}}\Phi^{\nu\mu})+
\frac{m^{2}}{h^{2}}\Phi^{\nu}+\added{\frac{i}{h}}e A_{\mu}\Phi^{\nu\mu}+
\added{\frac{i}{h}}eF^{\nu\mu}\Phi_{\mu}=0,\label{21}
\end{equation}
here $|\textbf{g}|$ is a coefficients matrix, $m$ is
particles mass and $\Phi^{\mu\nu}$ is anti-symmetric tensor, since
\begin{equation}
\Phi_{\nu\mu}=\partial_{\nu}\Phi_{\mu}-\partial_{\mu}\Phi_{\nu}+
\added{\frac{i}{h}}e A_{\nu}\Phi_{\mu}-\added{\frac{i}{h}}e A_{\mu}\Phi_{\nu}~~
\textmd{and}~~
F^{\mu\nu}=\triangle^{\mu}A^{\nu}-\triangle^{\nu}A^{\mu},\nonumber
\end{equation}
where $A_\mu$ is the vector potential of the charged black hole \added{and $A_{0}$ and $A_{3}$ are the components of $A_\mu$}, $e$
is the charge of the particle and $\triangle_{\mu}$ is covariant
derivative. The values of $\Phi^{\mu}$ and $\Phi^{\nu\mu}$ are given
by,
\begin{eqnarray*}
\Phi^{0}&=&\added{\frac{\Phi_{0}}{G}},~~~ \Phi^{1}=
\frac{\Phi_{1}}{B},~~~ \Phi^{2}
=\frac{\Phi_{2}}{C},~~~
\Phi^{3}=\frac{\Phi_{3}}{D},~~~
\Phi^{01}=\added{\frac{\Phi_{01}}{GB}},~~~\Phi^{02}=
\added{\frac{\Phi_{02}}{GC}},\\
\Phi^{03}&=&\added{\frac{\Phi_{03}}{GD}},~~~
\Phi^{12}=\frac{\Phi_{12}}{BC},
~~~\Phi^{13}=\frac{\Phi_{13}}{BD},~~~
\Phi^{23}=\frac{\Phi_{23}}{CD}.
\end{eqnarray*}
\added{Using WKB approximation for the  wave function ansatz:}
\cite{R3}, i.e.,
\begin{equation}
\Phi_{\nu}=c_{\nu}\exp\left[\added{\frac{i}{\hbar}}I_{0}(t,r,\theta,\phi)+
\sum_{\zeta=1}^{\zeta=n} \hbar^{\zeta}\added{I_{\zeta}}(t,r,\theta,\phi)\right],
\end{equation}
to the Lagrangian Eq.(\ref{21}) (where \added{$I_{0}$}  and $I_{i}$ are
corresponds to particles action, for $i=1,2,3,...$) and by
neglecting the higher order terms, we get the following set of
equations given below
\begin{eqnarray}
&&\frac{1}{B}[c_{1}(\partial_{0}\added{I_{0}})
(\partial_{1}\added{I_{0}})+e A_{0}c_{1}(\partial_{1}\added{I_{0}})
-c_{0}(\partial_{1}\added{I_{0}})^{2}]+\frac{1}{C}
[c_{2}(\partial_{0}\added{I_{0}})(\partial_{2}\added{I_{0}})-c_{0}\nonumber\\
&&(\partial_{2}\added{I_{0}})^{2}
+eA_{0}c_{2}(\partial_{2}\added{I_{0}})]+\frac{1}{D}
\left[c_{3}(\partial_{0}\added{I_{0}})(\partial_{3}\added{I_{0}})
+eA_{0}c_{3}(\partial_{3}\added{I_{0}})
-c_{0}(\partial_{3}\added{I_{0}})^{2}\right]\nonumber\\
&&+\frac{eA_{3}}{D}[c_{3}
(\partial_{0}\added{I_{0}})+eA_{0}c_{3}-c_{0}(\partial_{3}\added{I_{0}})]
-m^{2}c_{0}=0,\label{31}\\
&&\frac{1}{\added{G}(r)}[c_{0}(\partial_{0}\added{I_{0}})(\partial_{1}\added{I_{0}})-
eA_{0}c_{1}(\partial_{0}\added{I_{0}})-c_{1}(\partial_{0}\added{I_{0}})^{2}]-
\frac{1}{C}[c_{2}(\partial_{1}\added{I_{0}})
(\partial_{2}\added{I_{0}})-c_{1}\nonumber\\
&&(\partial_{2}\added{I_{0}})^{2}]+\frac{1}{D}[c_{3}
(\partial_{1}\added{I_{0}})
(\partial_{3}\added{I_{0}})-c_{1}(\partial_{3}\added{I_{0}})
-eA_{3}c_{1}(\partial_{3}\added{I_{0}})]
+\frac{eA_{3}}{D}[c_{3}((\partial_{1}\added{I_{0}})\nonumber\\
&&-eA_{3})c_{1}-c_{1}(\partial_{3}\added{I_{0}})]
-m^{2}c_{1}=0,~~\label{41}\end{eqnarray}\begin{eqnarray}
&&\frac{1}{\added{G}(r)}\left[c_{0}(\partial_{0}\added{I_{0}})
(\partial_{2}\added{I_{0}})-c_{2}(\partial_{0}\added{I_{0}})^{2}
-eA_{0}(\partial_{0}\added{I_{0}})c_{2}\right]
-\frac{1}{B}[c_{2}(\partial_{1}\added{I_{0}})^{2}
-c_{1}(\partial_{1}\added{I_{0}})\nonumber\\
&&(\partial_{2}\added{I_{0}})]-\frac{1}{D}
[c_{3}(\partial_{2}\added{I_{0}})(\partial_{3}\added{I_{0}})
-c_{2}(\partial_{3}\added{I_{0}})^{2}-eA_{3}c_{2}(\partial_{3}\added{I_{0}})]
+\frac{eA_{0}}{\added{G}(r)}[c_{0}(\partial_{2}\added{I_{0}})\nonumber\\&&-c_{2}
(\partial_{0}\added{I_{0}})
-c_{2}eA_{0}]+\frac{eA_{3}}{D}[c_{3}(\partial_{2}\added{I_{0}})
-c_{2}(\partial_{3}\added{I_{0}})
-eA_{3}c_{2}]-m^{2}c_{2}=0,\label{51}\\
&&\frac{1}{\added{G}(r)}[c_{0}(\partial_{0}\added{I_{0}})(\partial_{3}\added{I_{0}})
-c_{3}(\partial_{0}\added{I_{0}})^{2}+eA_{3}c_{0}(\partial_{0}\added{I_{0}})
-eA_{0}c_{3}(\partial_{0}\added{I_{0}})]
+\frac{1}{B}[c_{1}
(\partial_{1}\added{I_{0}})\nonumber\\
&&(\partial_{3}\added{I_{0}})-c_{3}(\partial_{1}\added{I_{0}})^{2}
+eA_{3}c_{1}(\partial_{1}\added{I_{0}})]+\frac{1}{C}
[c_{2}(\partial_{2}\added{I_{0}})(\partial_{3}\added{I_{0}})
-c_{3}
(\partial_{2}\added{I_{0}})^{2}
+eA_{3}c_{2}\nonumber\\
&&(\partial_{2}\added{I_{0}})]
+\frac{eA_{0}}{\added{G}(r)}[c_{0}(\partial_{3}\added{I_{0}})
-c_{3}(\partial_{0}\added{I_{0}})+eA_{3}c_{0}
-eA_{0}c_{3}]-m^{2}c_{3}=0.\label{61}
\end{eqnarray}
We can choose \added{$I_{0}$} by using separation of variables technique,
i.e.,
\begin{equation}
\added{I_{0}}=-(E-\tilde{\omega}\tilde{\Omega})t
+W(r)+\tilde{\omega}\phi+\upsilon(\theta),
\end{equation}
where \added{$\tilde{\Omega}$ is the angular momentum of the BH} and $E$ and $\tilde{\omega}$ represent particle energy and angular
momentum, respectively. From Eqs.(\ref{31})-(\ref{61}), we can
obtain the following matrix equation
\begin{equation*}
\tilde{K}(c_{0},c_{1},c_{2},c_{3})^{T}=0,
\end{equation*}
which provides ``$\tilde{K}$" as a matrix of order $4\times4$ and
its components are given by:
\begin{eqnarray}
\tilde{K}_{00}&=&-\frac{\dot{W}^{2}}{B}-\frac{\tilde{\omega}}{C}
-\frac{\dot{\upsilon}}{D}-\dot{\upsilon}eA_{3}
-m^{2},~~~
\tilde{K}_{01}=-\frac{\dot{W}(E-
\tilde{\omega}\tilde{\Omega})}{B}+\frac{\dot{W}eA_{0}}{B}\nonumber\\
\tilde{K}_{02}&=&\frac{\dot{\upsilon}
(E-\tilde{\omega}\tilde{\Omega})}{C},~~~
\tilde{K}_{03}=\frac{\dot{\upsilon}
(E-\tilde{\omega}\tilde{\Omega})}
{D}+eA_{0}\dot{\upsilon}+eA_{3}(E-
\tilde{\omega}\tilde{\Omega})+eA_{0},\nonumber\\
\tilde{K}_{10}&=&(E-\tilde{\omega}\tilde{\Omega})\frac{\dot{W}}{\added{G}(r)},\nonumber\\
\tilde{K}_{11}&=& \frac{-(E-\tilde{\omega}\tilde{\Omega})^{2}}{\added{G}(r)}-
\frac{(E-\tilde{\omega}\tilde{\Omega})eA_{0}}{\added{G}(r)}
-\frac{{\tilde{\omega}}^{2}}{C}-\frac{\dot{\upsilon}}{D}-
\frac{eA_{3}\dot{\upsilon}}{D}-eA_{3}\dot{\upsilon}-e^{2}A^{2}_{3}
\nonumber\\&-&m^{2},\nonumber\\
\tilde{K}_{12}&=&\frac{\dot{W}\tilde{\omega}}{C},~~
\tilde{K}_{13}=\frac{\dot{\upsilon\dot{W}}}{D}+eA_{3}\dot{W},
\tilde{K}_{20}=\frac{\dot{\upsilon}
(E-\tilde{\omega}\tilde{\Omega})}{A}+eA_{0}\frac{\dot{\upsilon}}{\added{G}(r)}
,\tilde{K}_{21}=-\dot{W}\dot{\upsilon},\nonumber
\end{eqnarray}
\begin{eqnarray}
\tilde{K}_{22}&=&-\frac{1}{\added{G}(r)}[(E-\tilde{\omega}\tilde{\Omega})^{2}+eA_{0}
(E-\tilde{\omega}\tilde{\Omega})]+\frac{1}{B}\dot{W}-\frac{1}{D}
[\dot{\upsilon}-eA_{3}\dot{\upsilon}]-
\frac{eA_{3}}{D}[\dot{\upsilon}+eA_{3}]\nonumber\\&-&m^{2},~~
\tilde{K}_{23}=\frac{1}{D}\tilde{\omega}\dot{\upsilon}+\frac{eA_{3}\tilde{\omega}}{D},~~
\tilde{K}_{30}=\frac{1}{\added{G}(r)}[(E-\tilde{\omega}\tilde{\Omega})\dot{\upsilon}
+eA_{3}(E-\tilde{\omega}\tilde{\Omega})]+\frac{eA_{0}}{\added{G}(r)}\nonumber\\
&&[\dot{\upsilon}+eA_{3}],~~~\tilde{K}_{31}=\frac{1}{B}
[\dot{W}\dot{\upsilon}+eA_{3}\dot{W}],~~~
\tilde{K}_{32}=\frac{1}{C}[\tilde{\omega}\dot{\upsilon}
+eA_{3}\dot{\upsilon}],\nonumber\\
\tilde{K}_{33}&=&-\frac{1}{\added{G}(r)}[(E-\tilde{\omega}\tilde{\Omega})^{2}+
(E-\tilde{\omega}\tilde{\Omega})eA_{0}]-
\frac{1}{B}\dot{W}^{2}-\frac{\dot{\upsilon}^{2}}{C}
-\frac{eA_{0}}{\added{G}(r)}[(E-\tilde{\omega}\tilde{\Omega})+eA_{0}]\nonumber\\
&&-m^{2},\nonumber
\end{eqnarray}
where $\dot{W}=\partial_{r}$\added{$I_{0}$},
$\dot{\upsilon}=\partial_{\theta}$\added{$I_{0}$}~and
$\tilde{\omega}=\partial_{\phi}$\added{$I_{0}$}.
For the non-trivial solution $\mid\tilde{\textbf{K}}\mid=0$ and
solving above equations one can yield
\begin{equation}\label{a1111}
ImW^{\pm}=\pm \int\sqrt{\frac{(E-eA_{0}-
\tilde{\omega}\tilde{\Omega}-eA_{3})^{2}+X}{\added{G}(r)B^{-1}}}dr,
\end{equation}
where $+$ and $-$ represent the outgoing and incoming particles,
respectively. Whereas, $``X"$ is the function which can be defined
as
\begin{equation}
X=-\frac{\added{G}(r)}{C}{\dot{\upsilon}}^{2}-m^{2}\added{G}(r)+2eA_{3}
(E-\tilde{\omega}\tilde{\Omega})+2e^{2}A_{0}A_{3}
-e^{2}A_{3}^{2}
\end{equation}
and $\tilde{\omega}$ is the angular
velocity at event horizon.

By integrating Eq.(\ref{a1111}) around the pole, we get
\begin{equation}
ImW^{\pm}
=\pm i\pi\frac{(E-eA_{0}-\tilde{\omega}\tilde{\Omega}-eA_{3})}{2\rho(r_{+})},
\end{equation}
where the surface gravity \added{$\kappa(r_{+})$~} of the charged black hole is
given by
\begin{equation}
\added{\kappa}(r_{+})=[\frac{2M}{r^{2}}-\frac{2Q^{2}}{r^{3}}
+\frac{1-6\kappa\lambda}{1-3\kappa\lambda}
N_{d}r^{\frac{3\kappa\lambda}{3\kappa\lambda-1}}]^{2}_{r=r_+}.
\end{equation}
The tunneling probability $\Gamma$ for outgoing charged vector
particles can be obtained by
\begin{eqnarray}\nonumber
\Gamma(ImW^{+})=&&\frac{\textmd{Prob{[emission]}}}{\textmd{Prob{[absorption]}}}=
\frac{\textmd{exp}[-2(ImW^++Im\Phi)]}
{\textmd{exp}[-2(ImW^{-}-Im\Phi)]}
={\textmd{exp}[-4ImW^+]}\\\nonumber =&&\exp\left[-
\frac{2\pi(E-eA_{0}-\tilde{\omega}\tilde{\Omega}
-eA_{3})}{\left(\frac{2M}{r^{2}}
-\frac{2Q^{2}}{r^{3}}+\frac{1-6\kappa\lambda}{1-3\kappa\lambda}
N_{d}r^{\frac{3\kappa\lambda}{3\kappa\lambda-1}}\right)^{2}}\right].
\end{eqnarray}
Now, we can calculate the $\tilde{T}_{H}(ImW^{+})$
by comparing the $\tilde{\Gamma}(ImW^{+})$ with the Boltzmann formula
$\tilde{\Gamma}_B(ImW^{+})\approx e^{-(E-
eA_{0}-\tilde{\omega}\tilde{\Omega}-eA_{3})/\tilde{T}_H(ImW^{+})}$, we get
\begin{equation}\label{aTH}
T_{H}(ImW^{+})=\left[\frac{M}{\pi r^{2}} -\frac{Q^{2}}{\pi
r^{3}}+\frac{1-6\kappa\lambda}{2\pi(1-3\kappa\lambda)}
N_{d}r^{\frac{3\kappa\lambda}{3\kappa\lambda-1}}\right]^{2}_{\added{r=r_+}~}.
\end{equation}
The result shows that the $\tilde{\Gamma}(ImW^{+})$ is depending on
$r$, the vector potential components ($A_{0}$ and $A_{3}$), energy
$E$, angular momentum $\tilde{\Omega}$, mass of black hole $M$,
$\kappa$ and $\lambda$ are the Rastall geometric parameters, $N_{d}$
and $Q$ dust field structure parameter and charge of black hole
respectively.

\subsection{The Black Hole Surrounded by the Radiation Field}

Second example is the line element of black hole surrounded by the radiation field
\cite{R1} is given below
\begin{eqnarray}
ds^{2}&=&-\left(1-\frac{2M}{r}+\frac{Q^{2}-N_{r}}{r^{2}}\right)dt^{2}
+\frac{dr^{2}}{\left(1-\frac{2M}{r}+\frac{Q^{2}-N_{r}}{r^{2}}\right)}\nonumber\\
&&+r^{2}(d\theta^{2}+\sin^{2}\theta d\phi^{2}),\label{a1}
\end{eqnarray}
where $M$ is a mass of black hole, $N_{r}$ is the negative radiation
structure parameters and $Q$ is a charge of black hole.

Following the procedure given in the preceding Section \textbf{2.1}
for this line element, we can obtain the surface gravity
\added{$\kappa(r_{+})$~} of this charged black hole surrounded by the
radiation field in the following form
\begin{equation}
\added{\kappa}(r_{+})=\left[\frac{2M}{r^{2}}-\frac{2Q^{2}}{r^{3}}
-\frac{r^{2}\dot{N}_{r}-2rN_{r}}{r^{4}}\right]^{2}_{r=r_+},
\end{equation}
where $\dot{N}_{r}=\frac{\partial N_{r}}{\partial r}$. The tunneling
rate of particles can be calculated as
\begin{equation}
\tilde{\Gamma}(ImW^{+})=\exp\left[-
\frac{4\pi(E-eA_{\mu}-\tilde{\omega}\tilde{\Omega})r^{6}}{\left[2Mr
-2Q^{2}-r\dot{N}_{r}+2N_{r}\right]^{2}}\right]
\end{equation}
and the corresponding Hawking temperature at horizon can be obtained
as
\begin{equation}
\tilde{T}(r_{+})=\frac{1}{4\pi r^{6}}\left[2Mr
-2Q^{2}-r\dot{N}_{r}+2N_{r}\right]^{2}|_{r=r_+}.
\end{equation}
This temperature depend on radiation structure parameter $N_{r}$,
mass $M$ and black hole charge $Q$.

\subsection{The Black Hole Surrounded by the Quintessence Field}

Third example is the line element of the black hole surrounded by the quintessence
field \cite{R1} is given below
\begin{eqnarray}
ds^{2}&=&-\left(1-\frac{2M}{r}+\frac{Q^{2}}{r^{2}}-
\frac{N_{q}}{r^{\frac{-1-2\kappa\lambda}{1-\kappa\lambda}}}\right)dt^{2}
+\frac{dr^{2}}{\left(1-\frac{2M}{r}+\frac{Q^{2}}{r^{2}}-
\frac{N_{q}}{r^{\frac{-1-2\kappa\lambda}{1-\kappa\lambda}}}\right)}\nonumber\\
&+&r^{2}(d\theta^{2}+\sin^{2}\theta d\phi^{2}),
\end{eqnarray}
where $N_{q}$ is a quintessence field structure parameter. By
following the same process and using the vector potential for this
black hole, the surface gravity can be derived as
\begin{equation}
\added{\kappa}(r_{+})=\left[\frac{2M}{r^{2}}-\frac{2Q^{2}}{r^{3}}
-\left(\frac{1+2\kappa\lambda}{1-\kappa\lambda}\right)
N_{q}r^{\frac{3\kappa\lambda}{1-\kappa\lambda}}\right]^{2}_{r=r_+}.
\end{equation}
The corresponding tunneling probability
\begin{equation}
\bar{\Gamma}=\exp\left[\frac{-4\pi(E-eA_{\mu}-\tilde{\omega}\tilde{\Omega})}
{\left[\frac{2M}{r^{2}}-\frac{2Q^{2}}{r^{3}}
+\left(\frac{1-6\kappa\lambda}{1-3\kappa\lambda}\right)
N_{d}r^{\frac{3\kappa\lambda}{3\kappa\lambda-1}}\right]^{2}}\right],
\end{equation}
and Hawking temperature
\begin{equation}
\bar{T}=\added{\left[\frac{\left(\frac{2M}{r^{2}}-\frac{2Q^{2}}{r^{3}}
-\left(\frac{1+2\kappa\lambda}{1-\kappa\lambda}\right)
N_{q}r^{\frac{3\kappa\lambda}{1-\kappa\lambda-}}\right)^{2}} {4\pi}\right]_{r=r_+}~}.
\end{equation}
are derived given in above expressions. The Hawking temperature
depends on $M$, $Q$ and $N_{q}$, i.e., quintessence field structure
parameter, mass
and charged of black hole, respectively.

\subsection{The Black Hole Surrounded by the Cosmological Constant Field}

Fourth example is the line element of black hole surrounded by the cosmological
constant field is given below \cite{R1}
\begin{eqnarray}
ds^{2}&=&-\left(1-\frac{2M}{r}+\frac{Q^{2}}{r^{2}}-
N_{c}r^{2}\right)dt^{2}
+\frac{dr^{2}}{\left(1-\frac{2M}{r}+\frac{Q^{2}}{r^{2}}-N_{c}r^{2}\right)}\nonumber\\
&+&r^{2}(d\theta^{2}+\sin^{2}\theta d\phi^{2}),
\end{eqnarray}
where $N_{c}$ is a cosmological constant field structure parameter.
For this black hole, the surface gravity at outer horizon is
obtained by following the above mentioned similar procedure, i.e.,
\begin{equation}
\added{\kappa}(r_{+})=\left[\frac{2M}{r^{2}}-\frac{2Q^{2}}{r^{3}}
-2N_{c}r\right]^{2}_{r=r_+}.
\end{equation}
Moreover, the required tunneling probability of particles
\begin{equation}
\hat{\Gamma}=\exp\left[\frac{-4\pi(E-eA_{\mu}-\tilde{\omega}\tilde{\Omega})}
{\left(\frac{2M}{r^{2}}-\frac{2Q^{2}}{r^{3}}
-2N_{c}r\right)^{2}}\right],
\end{equation}
and their corresponding Hawking temperature is calculated in the
following expression, i.e.,
\begin{equation}
\hat{T}=\added{\left[\frac{\left(\frac{2M}{r^{2}}-\frac{2Q^{2}}{r^{3}}
-2N_{c}r\right)^{2}} {4\pi}\right]_{r=r_+}~}.
\end{equation}
This temperature depend on $N_{c}$, $M$ and $Q$, i.e., cosmological
constant field structure parameter, mass
and charged of black hole, respectively.

\subsection{The Black Hole Surrounded by the Phantom Field}

Last example is the line element of black hole surrounded by the phantom field is
\cite{R1}
\begin{eqnarray}
ds^{2}&=&-\left(1-\frac{2M}{r}+\frac{Q^{2}}{r^{2}}-
\frac{N_{p}}{r^{\frac{-3+2\kappa\lambda}{1+\kappa\lambda}}}\right)dt^{2}
+\frac{dr^{2}}{\left(1-\frac{2M}{r}+\frac{Q^{2}}{r^{2}}-
\frac{N_{p}}{r^{\frac{-3+2\kappa\lambda}{1+\kappa\lambda}}}\right)}\nonumber\\
&+&r^{2}(d\theta^{2}+\sin^{2}\theta d\phi^{2}),
\end{eqnarray}
where $N_{p}$ is a phantom field structure parameter. For vector
potential $A_{\mu}$ of this black hole, the surface gravity can be
derived as
\begin{equation}
\added{\kappa}(r_{+})=\left[\frac{2M}{r^{2}}-\frac{2Q^{2}}{r^{3}}
-\left(\frac{3-2\kappa\lambda}{1+\kappa\lambda}\right)
N_{p}r^{\frac{2-3\kappa\lambda}{1+\kappa\lambda}}\right]^{2}.
\end{equation}
While, the tunneling probability of particles
\begin{equation}
\check{\Gamma}=\exp\left[\frac{-4\pi(E-eA_{\mu}
-\tilde{\omega}\tilde{\Omega})}{\left(\frac{2M}{r^{2}}-\frac{2Q^{2}}{r^{3}}
-\left(\frac{3-2\kappa\lambda}{1+\kappa\lambda}\right)
N_{p}r^{\frac{2-3\kappa\lambda}{1+\kappa\lambda}}\right)^{2}}\right]
\end{equation}
and the required Hawking temperature of particles can be obtained as
given below
\begin{equation}
\check{T}=\added{\left[\frac{\left(\frac{2M}{r^{2}}-\frac{2Q^{2}}{r^{3}}
-\left(\frac{3-2\kappa\lambda}{1+\kappa\lambda}\right)
N_{p}r^{\frac{2-3\kappa\lambda}{1+\kappa\lambda}}\right)^{2}}{4\pi}\right]_{r=r_+}~}
.
\end{equation}
The Hawking temperature depends on $M$, $Q$ and $N_{p}$, these are
phantom field structure parameter, mass
and charged of black hole, respectively.

\section{GUP-corrected Proca equation and the corrected Hawking Temperature}

In this section, we focus on the effect of the GUP on the tunneling of massive uncharged vector particles from the black hole solutions surrounded by perfect fluid in Rastall theory. Firstly we use the GUP-corrected Lagrangian for the massive uncharged vector field $\psi_{\mu}$ given by \cite{xiang2}
\begin{equation}
L_{GUP}=-\frac{1}{2}\left(D_{\mu}\psi_{\nu}-D_{\nu}\psi_{\mu}\right)\left(D^{\mu}\psi^{\nu}-D^{\nu}\psi^{\mu}\right)-\frac{m_W^2}{\hbar^2}\psi_{\mu}\psi^{\mu}.
\end{equation}

\added{One~} can derive the equation of the motion for the GUP-corrected lagrangian of massive uncharged vector field as follows:
\cite{xiang2}:
\begin{equation}
\partial_{\mu}\left(\sqrt{-g}\psi^{\mu\nu}\right)-\sqrt{-g}\frac{m_W^{2}}{\hbar^{2}}\psi^{\nu}+\beta\hbar^{2}\partial_{0}\partial_{0}\partial_{0}\left(\sqrt{-g}g^{00}\psi^{0\nu}\right)-\beta\hbar^{2}\partial_{i}\partial_{i}\partial_{i}\left(\sqrt{-g}g^{ii}\psi^{i\nu}\right)=0,\label{Fe2}
\end{equation}
with \begin{equation} \psi_{\mu\nu}=(1-\beta \hbar^2 \partial_{\mu}^2)\,\partial_{\mu}\psi_{\nu}-(1-\beta \hbar^2 \partial_{\nu}^2)\,\partial_{\nu}\psi_{\mu} .\end{equation}
It is noted that we use the  Latin indices for the modified tensor $\psi_{i\mu}$ as follows: $i=1,2,3$, on the other hand, for $\psi_{0\mu}$, we use the $0$ for the time coordinate. Moreover, we note that $\beta = 1/(3M_{f}^2 )$, which \added{$M_{f}$ is the Planck mass} and $m_W$ stands for the mass of the particle.

\subsection{The Black Hole Surrounded by the Dust Field in Rastall theory}
The metric of the is
given by 
\begin{equation}
ds^{2}=-\added{G}(r)(r)dt^{2}+B(r)dr^{2}+C(r)d\theta^{2}+D(r)d\phi^{2},\label{RNmetric}
\end{equation}
where $\added{G}(r)$, $B(r)$, $C(r)$ and $D(r)$ are given below:
\begin{eqnarray}
\added{G}(r)=1-\frac{2M}{r}+\frac{Q^{2}}{r^{2}}-
\frac{N_{d}}{r^{\frac{1-6\kappa\lambda}{1-3\kappa\lambda}}},~~
C(r)=r^{2},\\B(r)=\added{\frac{1}{G(r)}}=\frac{1}{1-\frac{2M}{r}+\frac{Q^{2}}{r^{2}}-
\frac{N_{d}}{r^{\frac{1-6\kappa\lambda}{1-3\kappa\lambda}}}},~~
D(r)=r^{2}\sin^2\theta.
\end{eqnarray}
Using the WKB method, we define the $\psi_{\mu}$ as follows:
\begin{equation}
\mathfrak{\varPsi}_{\mu}=c_{\mu}(t,r,\theta,\phi){\rm exp}\left[\added{\frac{i}{\hbar}}I(t,r,\theta,\phi)\right],\label{WKBW}
\end{equation}
where $I$ is defined as 
\begin{equation}
I(t,r,\theta,\phi)=I_{0}(t,r,\theta,\phi)+\hbar I_{1}(t,r,\theta,\phi)+\hbar^{2}I_{2}(t,r,\theta,\phi)+\cdots.\label{S0123}
\end{equation}
We use Eqs~\eqref{WKBW},~\eqref{S0123}, and the metric~\eqref{RNmetric} into Eq.~\eqref{Fe2}, then we only consider the lowest order terms in $\hbar$ to calculate the equations with
the corresponding coefficients $c_{\mu}$:
\begin{eqnarray}
\added{G}(r)\left[c_{0}(\partial_{r}I_{0})^{2}\mathcal{A}_{1}^{2}-c_{1}(\partial_{r}I_{0})(\partial_{t}I_{0})\mathcal{A}_{1}\mathcal{A}_{0}\right]\nonumber \\
+\frac{1}{C(r)}\left[c_{0}(\partial_{\theta}I_{0})^{2}\mathcal{A}_{2}^{2}-c_{2}(\partial_{\theta}I_{0})(\partial_{t}I_{0})\mathcal{A}_{2}\mathcal{A}_{0}\right]\nonumber \\
+\frac{1}{D(r)}\left[c_{0}(\partial_{\phi}I_{0})^{2}\mathcal{A}_{3}^{2}-c_{3}(\partial_{\phi}I_{0})(\partial_{t}I_{0})\mathcal{A}_{3}\mathcal{A}_{0}\right]+c_{0}m_{W}^{2}=0,
\end{eqnarray}
\begin{eqnarray}
-\frac{1}{\added{G}(r)}\left[c_{1}(\partial_{t}I_{0})^{2}\mathcal{A}_{0}^{2}-c_{0}(\partial_{t}I_{0})(\partial_{r}I_{0})\mathcal{A}_{0}\mathcal{A}_{1}\right]\nonumber \\
+\frac{1}{C(r)}\left[c1(\partial_{\theta}I_{0})^{2}\mathcal{A}_{2}^{2}-c_{2}(\partial_{\theta}I_{0})(\partial_{r}I_{0})\mathcal{A}_{2}\mathcal{A}_{1}\right]\nonumber \\
+\frac{1}{D(r)}\left[c_{1}(\partial_{\phi}I_{0})^{2}\mathcal{A}_{3}^{2}-c_{3}(\partial_{\phi}I_{0})(\partial_{r}I_{0})A_{3}\mathcal{A}_{1}\right]+c_{1}m_{W}^{2}=0,
\end{eqnarray}
\begin{eqnarray}
-\frac{1}{\added{G}(r)}\left[c_{2}(\partial_{t}I_{0})^{2}\mathcal{A}_{0}^{2}-c_{0}(\partial_{t}I_{0})(\partial_{\theta}I_{0})\mathcal{A}_{0}\mathcal{A}_{2}\right]\nonumber \\
+\added{G}(r)\left[c_{2}(\partial_{r}I_{0})^{2}\mathcal{A}_{1}^{2}-c_{1}(\partial_{r}I_{0})(\partial_{\theta}I_{0})\mathcal{A}_{1}\mathcal{A}_{2}\right]\nonumber \\
+\frac{1}{D(r)}\left[c_{2}(\partial_{\phi}I_{0})^{2}\mathcal{A}_{3}^{2}-c_{3}(\partial_{\phi}I_{0})(\partial_{\theta}I_{0})\mathcal{A}_{3}\mathcal{A}_{2}\right]+c_{2}m_{W}^{2}=0,
\end{eqnarray}
\begin{eqnarray}
-\frac{1}{\added{G}(r)}\left[c_{3}(\partial_{t}I_{0})^{2}\mathcal{A}_{0}^{2}-c_{0}(\partial_{t}I_{0})(\partial_{\phi}I_{0})\mathcal{A}_{0}\mathcal{A}_{3}\right]\nonumber \\
+\added{G}(r)\left[c_{3}(\partial_{r}I_{0})^{2}\mathcal{A}_{1}^{2}-c_{1}(\partial_{r}I_{0})(\partial_{\phi}I_{0})\mathcal{A}_{1}\mathcal{A}_{3}\right]\nonumber \\
+\frac{1}{C(r)}\left[c_{3}(\partial_{\theta}I_{0})^{2}\mathcal{A}_{2}^{2}-c_{2}(\partial_{\theta}I_{0})(\partial_{\phi}I_{0})\mathcal{A}_{2}\mathcal{A}_{3}\right]+c_{3}m_{W}^{2}=0,
\end{eqnarray}
where the $\mathcal{A}_{\mu}$s are defined as 
\begin{eqnarray}
 &  & \mathcal{A}_{0}=1+\beta\added{\frac{1}{G(r)}}(\partial_{t}S_{0})^{2},\ \mathcal{A}_{1}=1+\beta\added{G}(r)(\partial_{r}S_{0})^{2},\nonumber \\
 &  & \mathcal{A}_{2}=1+\beta\frac{1}{C(r)}(\partial_{\theta}S_{0})^{2},\ \mathcal{A}_{3}=1+\beta\frac{1}{D(r)}(\partial_{\phi}S_{0})^{2}.
\end{eqnarray}
Using the semi-classical Hamilton-Jacobi method with WKB ansatz, we separate
the variables as follows:
\begin{equation}
I_{0}=-Et+R(r)+\Theta(\theta,\phi)+k,\label{RNs0fj}.
\end{equation}
Note that the energy of the radiated particle is defined with $E$. Afterwards, we obtain a matrix equation as follows:
\begin{equation}
\added{\cup}(c_{0},c_{1},c_{2},c_{3})^{T}=0,\label{matrixeq}
\end{equation}
where \added{$\cup$}  is a 4$\times$4 matrix, the elements of which are
\begin{eqnarray}
 && \added{\cup_{11}}=\added{G}(r)R'^{2}\mathcal{A}_{1}^{2}+\frac{{J_{\theta}}^{2}}{C(r)}\mathcal{A}_{2}^{2}+\frac{{J_{\phi}}^{2}}{D(r)}\mathcal{A}_{3}^{2}+m_{W}^{2},\ \added{\cup_{12}}=-\added{G}(r)R'(-E)\mathcal{A}_{1}\mathcal{A}_{0},\nonumber \\
&& \added{\cup_{13}}=-\frac{J_{\theta}(-E)}{C(r)}\mathcal{A}_{2}\mathcal{A}_{0},\ \added{\cup_{14}}=-\frac{J_{\phi}(-E)}{D(r)}\mathcal{A}_{3}\mathcal{A}_{0},\nonumber \\
&& \added{\cup_{21}}=\frac{(-E)R'}{\added{G}(r)}\mathcal{A}_{0}\mathcal{A}_{1},\ \added{\cup_{22}}=-\frac{(-E)^{2}}{\added{G}(r)}\mathcal{A}_{0}^{2}+\frac{{J_{\theta}}^{2}}{C(r)}\mathcal{A}_{2}^{2}+\frac{J_{\phi}^{2}}{D(r)}\mathcal{A}_{3}^{2}+m_{W}^{2},\nonumber \\
 && \added{\cup_{23}}=-\frac{J_{\theta}R'}{C(r)}\mathcal{A}_{2}\mathcal{A}_{1},\ \added{\cup_{24}}=-\frac{J_{\phi}R'}{D(r)}\mathcal{A}_{3}\mathcal{A}_{1},\nonumber \\
 && \added{\cup_{31}}=\frac{-EJ_{\theta}}{\added{G}(r)}\mathcal{A}_{0}\mathcal{A}_{2},\ \added{\cup_{32}}=-\added{G}(r)R'J_{\theta}\mathcal{A}_{1}\mathcal{A}_{2},\\
 & & \added{\cup_{33}}=-\frac{(-E)^{2}}{\added{G}(r)}\mathcal{A}_{0}^{2}+\added{G}(r)R'^{2}\mathcal{A}_{1}^{2}+\frac{J_{\phi}^{2}}{D(r)}\mathcal{A}_{3}^{2}+m_{W}^{2},\ \added{\cup_{34}}=-\frac{J_{\theta}J_{\phi}}{D(r)}\mathcal{A}_{3}\mathcal{A}_{2},\nonumber \\
 &  & \added{\cup_{41}}=\frac{(-E)J_{\phi}}{\added{G}(r)}\mathcal{A}_{0}\mathcal{A}_{3},\ \added{\cup_{42}}=-\added{G}(r)R'J_{\phi}\mathcal{A}_{1}\mathcal{A}_{3},\nonumber \\
 &  & \added{\cup_{43}}=-\frac{J_{\theta}J_{\phi}}{C(r)}\mathcal{A}_{2}\mathcal{A}_{3},\ \added{\cup_{44}}=-\frac{(-E)^{2}}{\added{G}(r)}\mathcal{A}_{0}^{2}+\added{G}(r)R'^{2}A_{1}^{2}+\frac{J_{\theta}^{2}}{C(r)}\mathcal{A}_{2}^{2}+m_{W}^{2},\nonumber 
\end{eqnarray}
where $R'=\partial_{r}R$, $J_{\theta}=\partial_{\theta}\Theta$ and
$J_{\phi}=\partial_{\phi}\Theta$.

It is noted that for the condition of ${\rm det}\cup=0$, we find the  nontrivial solution of the Eq.~\eqref{matrixeq}. First, we consider only the lowest order terms of $\beta$, then calculate the ${\rm det}\cup=0$. Our main aim is to obtain the radial part of the equation so that we integrate it using the complex integral method around the event horizon as follows:
\begin{equation}
\text{Im} R_{\pm}(r)=\pm \text{Im} \int dr \sqrt{-\frac{m^{2}}{\added{G}(r)}+\frac{E^{2}}{\added{G}(r)^{2}}-\frac{J_{\theta}^{2}+J_{\phi}^{2}}{\added{G}(r)D(r)}}\left(1+\frac{\mathcal{T}_{1}}{\mathcal{T}_{2}}\beta\right), \label{in1}
\end{equation}
where 
\begin{eqnarray}
\mathcal{T}_{1} & = & -3\added{G}(r)m^{4}C(r)+6m^{2}C(r)(E)^{2}-6\added{G}(r)m^{2}(J_{\theta}^{2}+\frac{J_{\phi}^{2}}{D(r)})-\frac{6\added{G}(r)J_{\theta}^{4}}{C(r)}\nonumber \\
 & + & 6(E)^{2}(J_{\theta}^{2}+\frac{J_{\phi}^{2}}{D(r)})-\frac{7\added{G}(r)J_{\theta}^{2}J_{\phi}^{2}}{D(r)}-\frac{3\added{G}(r)J_{\theta}^{4}J_{\phi}^{2}}{2m^{2}D(r)^{2}}-\frac{5\added{G}(r)J_{\phi}^{4}{\rm csc}^{4}\theta}{C(r)}+\frac{3\added{G}(r)J_{\theta}^{2}J_{\phi}^{4}}{2m^{2}D(r)^{2}},\\
\mathcal{T}_{2} & = & -\added{G}(r)m^{2}r^{2}+r^{2}(E)^{2}-\added{G}(r)(J_{\theta}^{2}+\frac{J_{\phi}^{2}}{D(r)}).
\end{eqnarray}

We rewrite the metric close to the event horizon to obtain solution for the integral:
\begin{equation}
\added{G}(r_h)\approx \frac{\Delta_{,r}  \,(r_{h})}{r_h{^2}}(r-r_h).
\end{equation}

Afterwards we manage to obtain the solution of the integral \eqref{in1} for the radial part as follows:
\begin{equation}
 \text{Im} R_{\pm}(r) = \pm i \pi\frac{r_{h}^{2}}{\Delta_{,r} \, (r_{h})}(E)\times\left(1+\beta\Xi\right),
\end{equation}
where $\Xi=6m^{2}+\frac{6}{r_{h}^{2}}\left(J_{\theta}^{2}+J_{\phi}^{2}{\rm csc}^{2}\theta\right)$.

It is quite clear that $\Xi>0$. We note that $R_{+}$ represents
the radial function for the outgoing particles and $R_{-}$ is for
the ingoing particles. Thus, the tunneling rate of $W$ bosons
near the event horizon is 
\begin{eqnarray}
\Gamma & = & \frac{P_{outgoing}}{P_{ingoing}}=\frac{{\rm exp}\left[-\frac{2}{\hbar}({\rm Im}R_{+}+{\rm Im}\Theta)\right]}{{\rm exp}\left[-\frac{2}{\hbar}({\rm Im}R_{-}+{\rm Im}\Theta)\right]}={\rm exp}\left[-\frac{4}{\hbar}{\rm Im}R_{+}\right]\nonumber \\
 & = & \exp\left[-\frac{4\pi}{\hbar}\frac{r_{h}^{2}}{\Delta_{,r}\, (r_{h})}(E)\times\left(1+\beta\Xi\right)\right].
\end{eqnarray}
If we set $\hbar=1$, we find the corrected Hawking temperature as follows: 
\begin{eqnarray}
T_{e-H}=\frac{\Delta_{,r} \, (r_{h})}{4\pi r_{h}^{2}\left(1+\beta\Xi\right)}=T_{0}\left(1-\beta\Xi\right),\label{effectiveHT}
\end{eqnarray}
where $T_{0}=\left[\frac{M}{\pi r^{2}} -\frac{Q^{2}}{\pi
r^{3}}+\frac{1-6\kappa\lambda}{2\pi(1-3\kappa\lambda)}
N_{d}r^{\frac{3\kappa\lambda}{3\kappa\lambda-1}}\right]^{2}${\added{$_{r=r_+}$}~is the original
Hawking temperature of a  corresponding black hole. We find the corrected Hawking temperature with the effect of quantum gravity. In addition, the Hawking temperature is increased if one use the quantum gravity effects, but then this effects are cancels in some point and black hole remnants are occurred.

\subsection{The Black Hole Surrounded by the Radiation Field}

Following the procedure given in the preceding Section \textbf{2.1}
for this line element, we obtain the corrected Hawking temperature with the effect of quantum gravity for this charged black hole surrounded by the
radiation field in the following form
\begin{eqnarray}
T_{e-H}=\frac{\Delta_{,r} \, (r_{h})}{4\pi r_{h}^{2}\left(1+\beta\Xi\right)}=T_{0}\left(1-\beta\Xi\right),\label{effectiveHT}
\end{eqnarray}
where original Hawking temperature is
\begin{equation}
T_{0}=\frac{1}{4\pi r^{6}}\left[2Mr
-2Q^{2}-r\dot{N}_{r}+2N_{r}\right]^{2}|_{r=r_+}.
\end{equation}
This temperature depend on radiation structure parameter $N_{r}$,
mass $M$ and black hole charge $Q$.
Moreover, the quantum effects explicitly counteract the temperature
increases during evaporation, which will cancels it out at some point.
Naturally, black hole remnants will be left.

\subsection{The Black Hole Surrounded by the Quintessence Field}

By following the same process, we calculate the corrected Hawking temperature under the effect of quantum gravity as follows:
\begin{eqnarray}
T_{e-H}=\frac{\Delta_{,r} \, (r_{h})}{4\pi r_{h}^{2}\left(1+\beta\Xi\right)}=T_{0}\left(1-\beta\Xi\right),\label{effectiveHT}
\end{eqnarray}
with the original Hawking temperature
\begin{equation}
T_{0}=\added{\left[\frac{\left(\frac{2M}{r^{2}}-\frac{2Q^{2}}{r^{3}}
-\left(\frac{1+2\kappa\lambda}{1-\kappa\lambda}\right)
N_{q}r^{\frac{3\kappa\lambda}{1-\kappa\lambda-}}\right)^{2}} {4\pi}\right]_{r=r_+}~},
\end{equation}
are derived given in above expressions. The Hawking temperature
depends on $M$, $Q$ and $N_{q}$, i.e., quintessence field structure
parameter, mass
and charged of black hole, respectively. Naturally, black hole remnants will be left.

\subsection{The Black Hole Surrounded by the Cosmological Constant Field}
One can repeat the process just for this black hole to calculate the corresponding corrected Hawking temperature with the quantum gravity effects as follows:
\begin{eqnarray}
T_{e-H}=\frac{\Delta_{,r} \, (r_{h})}{4\pi r_{h}^{2}\left(1+\beta\Xi\right)}=T_{0}\left(1-\beta\Xi\right),\label{effectiveHT}
\end{eqnarray}
with
\begin{equation}
T_{0}=\added{\left[\frac{\left(\frac{2M}{r^{2}}-\frac{2Q^{2}}{r^{3}}
-2N_{c}r\right)^{2}} {4\pi}\right]_{r=r_+}~}.
\end{equation}
This temperature depend on $N_{c}$, $M$ and $Q$, i.e., cosmological
constant field structure parameter, mass
and charged of black hole, respectively. Again here, remnants are left.

\subsection{The Black Hole Surrounded by the Phantom Field}

Last example is the line element of black hole surrounded by the phantom field. Now we again repeat the same process to obtain the following corrected Hawking temperature 
\begin{eqnarray}
T_{e-H}=\frac{\Delta_{,r} \, (r_{h})}{4\pi r_{h}^{2}\left(1+\beta\Xi\right)}=T_{0}\left(1-\beta\Xi\right),\label{effectiveHT}
\end{eqnarray} with
\begin{equation}
\check{T}=\added{\left[\frac{\left(\frac{2M}{r^{2}}-\frac{2Q^{2}}{r^{3}}
-\left(\frac{3-2\kappa\lambda}{1+\kappa\lambda}\right)
N_{p}r^{\frac{2-3\kappa\lambda}{1+\kappa\lambda}}\right)^{2}}{4\pi}\right]_{r=r_+}~}
.
\end{equation}
The Hawking temperature depends on $M$, $Q$ and $N_{p}$, these are phantom field structure parameter, mass
and charged of black hole, respectively.

\section{Conclusions}

In this research paper, we have successfully analyzed the GUP
corrected Hawking temperature of $W^{\pm}$ boson vector particles using the equation of motion for the Glashow-Weinberg-Salam model.
First of all, we analyzed the modified Hamilton-Jacobi equation
by resolving the modified Lagrangian equation utilized to the
magnetized particles in the spacetime. We have analyzed the GUP
effect on the radiation of black holes surrounded by perfect 
fluid in Rastall theory.

As the original Hawing radiation, the Hawking temperature $T_{H}$
of the black holes are associated to its mass $M$ and charged $Q$.
However, these results indicated that if the effect of quantum
gravity is counted and the behavior of the tunneling boson vector
particle on the event horizon is observe from the original event.
The Hawking temperature $T_{H}$ and tunneling probability $\Gamma$
quantities are not just sensitively dependent on the mass $M$ and
charged $Q$ of the black hole. The Hawking temperature $T_{H}$, tunneling
 probability $\Gamma$ and surface gravity $\kappa$ are only dependent on
 the geometry (structure parameter) of black hole. Moreover, the corrected Hawking temperature $T_{e-H}=T_{0}\left(1-\beta\Xi\right)$ have been calculated with the effect of quantum gravity. The Hawking temperature is increased if one use the quantum gravity effects, but then this effects are cancels in some point and black hole remnants are occurred.

\acknowledgements
This work is supported by Comisi\'on Nacional
de Ciencias y Tecnolog\'ia of Chile (CONICYT) through FONDECYT Grant N$^{\textup{o}}$ 3170035 (A. \"{O}.).  The authors declares that there is no conflict of interest regarding the publication of this paper.

\end{document}